\documentclass[12pt]{article}
\usepackage{graphicx}
 
\newcommand{\be}{\begin{equation}}
\newcommand{\ee}{\end{equation}}
\newcommand{\bea}{\begin{eqnarray}}
\newcommand{\eea}{\end{eqnarray}}
\newcommand{\nn}{\nonumber}

\newcommand{\psl}{p \kern-.45em{/}}
\newcommand{\qsl}{q \kern-.5em{/}}

\begin{document}
 
\title{Dilepton decays of baryon resonances} 
\author{M.\ Z\'et\'enyi\thanks{E-mail address: {\texttt zetenyi@rmki.kfki.hu}}
\enskip and \enskip Gy.\ Wolf\thanks{E-mail address: {\texttt zetenyi@rmki.kfki.hu}}\\
KFKI Research Institute for Nuclear and
Particle Physics, \\ 
 H-1525 Budapest 114, POB.\ 49, Hungary\\
}
 
\maketitle

\begin{abstract}
\noindent
Dalitz decay of baryon resonances is studied and 
expressions for the decay width are derived for resonances with 
arbitrary spin and parity. Contributions of the various terms in the 
transition matrix element are compared and relevance of spin-parity
and the resonance mass is discussed. Explicite algebraic expressions
are cited for spin$\le$5/2 resonances. The results can be used in models
of dielectron production in elementary reactions and heavy ion collisions.
\end{abstract}

\noindent
{\it keywords:} dileptons, baryon resonances, electromagnetic transitions

\section{Introduction}
The aim of relativistic heavy ion collision experiments is to study
the properties of the hot and dense nuclear matter created in the collisions.
These properties have to be reconstructed using the information encoded
in the particles detected in the final stage of the collision.
The best probes for this investigation are dileptons (in the few GeV 
per nucleon bombarding energy range dielectrons), 
which leave the hot and dense zone  unaffected by final state interactions.

Theoretical studies predict a modification of hadron properties
(masses and decay widths) in hot and dense nuclear matter.
These medium effects can most easily be studied in the case of vector mesons,
which decay directly to dileptons and their masses are measured through
the invariant mass of the resulting dilepton.

Dielectron production in heavy ion collisions with a few GeV per nucleon
energy has been studied experimentally by the DLS collaboration at 
BEVALAC and will be studied by the HADES at GSI.
In order to obtain a reliable model of heavy ion collisions first one 
has to check the elementary reaction channels used in the model.
For this purpose dilepton production in nucleon-nucleon collisions
has to be investigated both experimentally and theoretically.

Particle production in nucleon-nucleon collisions is usually described 
using the resonance model, i.e., the final state particles are
produced via the creation and subsequent decay of baryon resonances. 
This multistep process is factorized so that initial and final state
particles of the individual steps are handled as on-shell particles
and in case of resonances
their masses are generated according to a Breit-Wigner distribution.
Different factorizations result in different models with different elementary
channels.

In models of dielectron production in $pp$ collisions usually the
following elementary channels are included:
direct dileptonic decay of $\rho^0$, $\omega$ and $\phi$ vector mesons,
Dalitz decay of mesons $\pi^0\to\gamma e^+e^-$, $\eta\to\gamma e^+e^-$,
$\omega\to\pi^0 e^+e^-$ and Dalitz decay of the $\Delta(1232)$ resonance,
$\Delta\to Ne^+e^-$. Such models (e.g.\ \cite{giessen1,giessen2,frankfurt})
can reasonably well reproduce the experimental dilepton invariant mass spectra
in the $E_b$ = 1 -- 2 GeV bombarding energy range \cite{DLSpp}. 
In models of heavy ion collisions the additional channels of pion 
annihilation and $pn$ bremsstrahlung appear.

With increasing beam energy creation of higher mass baryon resonances
becomes more and more relevant. In heavy ion collisions kinetic energy
of colliding particles can be accumulated in multiple scattering processes,
therefore higher mass resonances appear at lower beam energies.
These resonances may contribute to the 
dilepton spectrum through their Dalitz decay similarly to the $\Delta(1232)$.
This contribution may especially be relevant in the higher dilepton mass
range where the $\rho^0$ and $\Delta(1232)$ channels are already negligible.
This dilepton mass range coincides with the range of the $\phi$ meson mass.

Recently, production of $\phi$ mesons in subthreshold heavy ion collisions has 
received some attention both experimentally \cite{Kotte} and 
theoretically \cite{Chung,ourpaper}. $\phi$ mesons are being studied
by the FOPI group at GSI via their kaon decay channel, $\phi\to K^+K^-$,
but the dileptonic decay channel will also be accessible by the HADES
detector. In the latter case an important source of background may come
from the Dalitz decay of higher mass resonances.

In this paper we derive expressions for the differential Dalitz decay
width of baryon resonances with arbitrary spin-parities. We discuss
in detail the Dalitz decay of resonances with spin$\le$5/2.
Our paper is organized as follows: in Sec.\ \ref{dal.dec.} we
derive expressions for the electromagnetic transition matrix elements of
baryon resonances with arbitrary spin-parity and give expressions for the 
Dalitz decay width. In Sec.\ \ref{results} we give a detailed discussion 
of the numerical results. We compare contributions of the various terms 
in the transition matrix element and discuss the relevance of spin-parity
and the resonance mass. In the Appendix we cite the spin projectors
used in the calculations and the explicite expressions for the
polarization averaged squared transition matrix elements of the
baryon resonance Dalitz decays.

%%%%%%%%%%%%%%%%%%%%%%%%%%%%%%%%%%%%%%%%%%%%%%%%%%%%%%%%%%%%%%%%%%%%%%%%
 
\section{Dalitz-decay of a spin-$J$ resonance}
\label{dal.dec.}

\subsection{Matrix elements of the electromagnetic current}
To study the electromagnetic decays of a spin-$J$ baryon resonance 
one has to determine the matrix element of the electromagnetic current 
operator $J_{\mu}$ between the nucleon ($N$) and the spin-$J$ baryon 
($R$) state $\langle N | J_{\mu} | R \rangle$.
Let $p_*$, $m_*$, and $\lambda_*$ denote the four-momentum, mass,
and helicity of the $R$ resonance, respectively, and $p$, $m$, and
$\lambda$ the corresponding quantities for the nucleon. Let us
introduce the notations $q = p_* - p$ and $P = (p_* + p)/2$.

For $J \ge 3/2$ the spin-$J$ fermion can be described by a generalized
Rarita-Schwinger spinor-tensor field
$\Psi^{\rho_1 \cdots \rho_n}$ ($n = J - 1/2$), therefore the
momentum-space wavefunction of the particle is
the spinor-tensor amplitude $u^{\rho_1 \cdots \rho_n}(p_*,\lambda_*)$.
This quantity in general contains some additional lower spin components.
The spin-$J$ content of $u^{\rho_1 \cdots \rho_n}(p_*,\lambda_*)$ 
can be selected by prescribing
the generalized Rarita-Schwinger relations:
\bea
u^{\cdots \rho_i \cdots \rho_k \cdots}(p_*,\lambda_*) & = &
   u^{\cdots \rho_k \cdots \rho_i \cdots}(p_*,\lambda_*), \nn \\
{u^{\cdots \sigma \cdots}}_{\sigma}{}^{\cdots}(p_*,\lambda_*) &
   = & 0, \nn \\ 
u^{\cdots \sigma \cdots}(p_*,\lambda_*) p_{*\sigma} & = & 0, \nn \\
u^{\cdots \sigma \cdots}(p_*,\lambda_*) \gamma_{\sigma} & = & 0.
\label{RSrel}
\eea
The matrix element of the electromagnetic current can be written generally as
\be
\langle N | J_{\mu} | R \rangle =
\bar{u}(p,\lambda) \Gamma_{\mu \rho_1 \cdots \rho_n}
u^{\rho_1 \cdots \rho_n}(p_*,\lambda_*),
\label{mxelement}
\ee
where the form of $\Gamma_{\mu \rho_1 \cdots \rho_n}$ is restricted
by the conservation of electric charge,
$q^{\mu} \langle N | J_{\mu} | R \rangle = 0$.
The relations (\ref{RSrel}) and the Dirac equations 
$\bar{u}(p,\lambda)(\psl - m) = 0$
and $(\psl_* - m_*) u^{\rho_1 \cdots \rho_n}(p_*,\lambda_*)
= 0$ reduce further the number of independent terms
in $\Gamma_{\mu \rho_1 \cdots \rho_n}$ to three, 
which can be chosen of the form
\be
\Gamma_{\mu \rho_1 \cdots \rho_n} =
\sum_{i=1}^{3} f_i(q^2) \chi^i_{\mu \rho_1}
p_{\rho_2} \cdots p_{\rho_n} G,
\label{gamma}
\ee
with
\bea
\chi^1_{\mu \rho} & = &
\gamma_{\mu} q_{\rho} - \qsl g_{\mu \rho}, \nn \\
\chi^2_{\mu \rho} & = &
P_{\mu} q_{\rho} - (P \cdot q) g_{\mu \rho}, \nn \\
\chi^3_{\mu \rho} & = &
q_{\mu} q_{\rho} - q^2 g_{\mu \rho},
\eea
and $G=1$ or $\gamma_5$ for resonances with positive or negative
normalities, respectively. The normality of a spin-$J$ baryon is
defined by $P (-1)^{J-1/2}$ with $P$ the intrinsic parity.
$f_i(q^2)$, $i = 1,2,3$ are three independent form factors.

In the spin-1/2 case the general form of the matrix element of the
electromagnetic current is
\be
\langle N | J_{\mu} | R \rangle =
\bar{u}(p,\lambda) \Gamma_{\mu}
u(p_*,\lambda_*),
\label{mxelement1/2}
\ee
and there are two independent terms in $\Gamma_{\mu}$:
\be
\Gamma_{\mu} =
\sum_{i=1}^{2} f_i(q^2) \chi^i_{\mu} G,
\label{gamma1/2}
\ee
with
\bea
\chi^1_{\mu} & = & (P\cdot q) \gamma_{\mu} - \qsl P_{\mu}, \nn \\
\chi^2_{\mu} & = & q^2 \gamma_{\mu} - \qsl q_{\mu}.
\eea
$G$ is again 1 or $\gamma_5$ depending on the normality of the resonance.

%-----------------------------------------------------------------------

\subsection{The Dalitz-decay width}
The differential width of the Dalitz-decay of a
particle is related to its photonic decay width to a
virtual photon, $\Gamma_{R \to N\gamma}(M)$, by \cite{LS}
\be
\frac{d\Gamma_{R \to N e^+ e^-}}{dM^2} =
\frac{\alpha}{3\pi}\frac{1}{M^2}\Gamma_{R \to N\gamma}(M).
\label{dalwid}
\ee
Here the notation $M^2 (=q^2)$ is used for the square of the
dilepton invariant mass (= mass of the virtual photon).
$\Gamma_{R \to N\gamma}(M)$ can be expressed in terms of the
photonic decay matrix element $\langle N \gamma | T | R \rangle$ as
\be
\Gamma_{R \to N\gamma}(M) =
\frac{\sqrt{\lambda (m_*^2,m^2,M^2)}}{16 \pi m_*^3}
\frac{1}{n_{pol,R}}
\sum_{pol} {\vert \langle N \gamma | T | R \rangle \vert}^2,
\label{photwid}
\ee
where the transition matrix element is related to the electromagnetic current
matrix element by
\be
\langle N \gamma | T | R \rangle =
- \epsilon^{\mu} \langle N | J_{\mu} | R \rangle,
\ee
with $\epsilon^{\mu}$ the photon polarization vector.
In (\ref{photwid}) $n_{pol,R}$ is the number of polarization states
of the $R$ resonance and
$\lambda(a,b,c) = a^2 + b^2 + c^2 - 2(ab+bc+ac)$ is the usual
kinematical factor.

The polarization sum in (\ref{photwid}) runs over all {\em physical}
polarization states of the incoming and outgoing particles, that is,
in the case of the spin-$J$ resonance only over those states that
belong to the spin-$J$ content of $u^{\rho_1 \cdots \rho_n}(p_*,\lambda_*)$.
As a result, the spin-$J$ projector appears in (\ref{photwid}) after
substituting (\ref{mxelement}). The projectors for spin-3/2 and 5/2
are given in Appendix A.

If the differential cross section of the production of the resonance R
in an elementary reaction, $d\sigma_R/dm_*$, is known Eq.\ (\ref{dalwid})
can be used to obtain the cross section of dilepton production
through the Dalitz decay of $R$ in the same reaction. This differential
cross section is given by
\be
\frac{d\sigma_{e^+e^-,R_{Dalitz}}}{dM} =
\int dm_* \frac{d\sigma_R}{dm_*}\frac{1}{\Gamma^R_{tot}(m_*)}
\frac{d\Gamma_{R \to N e^+ e^-}(m_*)}{dM}.
\ee

%-----------------------------------------------------------------------
 
\subsection{Form factors and coupling constants}

A possibility to determine the functional form of the form factors
$f_i(M^2)$ is the application of the vector dominance model (VDM)
\cite{Titov}.
However, we aim at a model that handles the theoretically more 
interesting vector meson direct dilepton decays separately from 
the baryon resonance Dalitz decays serving as a background.
Applying the VDM to the baryon resonance Dalitz decay channels
simultaneously with the inclusion of the vector meson direct decays
would give rise to a double counting if -- following the usual
treatment -- the model incorporates
vector mesons originating from decays of baryon resonances.
The authors of \cite{Tubingen} use a sophisticated method to 
avoid this double counting. In the present paper we do not
apply the VDM and keep the form factors $f_i(q^2)$ constant.

We introduce the dimensionless coupling constants $g_i$ via the equation
$f_i = e g_i / m^{k_i}$, where $e$ is the elementary charge related to the 
fine structure constant by $e^2 = 4\pi\alpha$. The exponent $k_i$ of the 
nucleon mass is determined by the requirement that the couplings
$g_i$ are dimensionless. The values of $g_i$ are obtained using
the real photonic decay width of the resonances.

%%%%%%%%%%%%%%%%%%%%%%%%%%%%%%%%%%%%%%%%%%%%%%%%%%%%%%%%%%%%%%%%%%%%%%%%

\section{Discussion of the results}
\label{results}

\begin{figure}[p]
\begin{center}
\includegraphics[width=6.2cm]{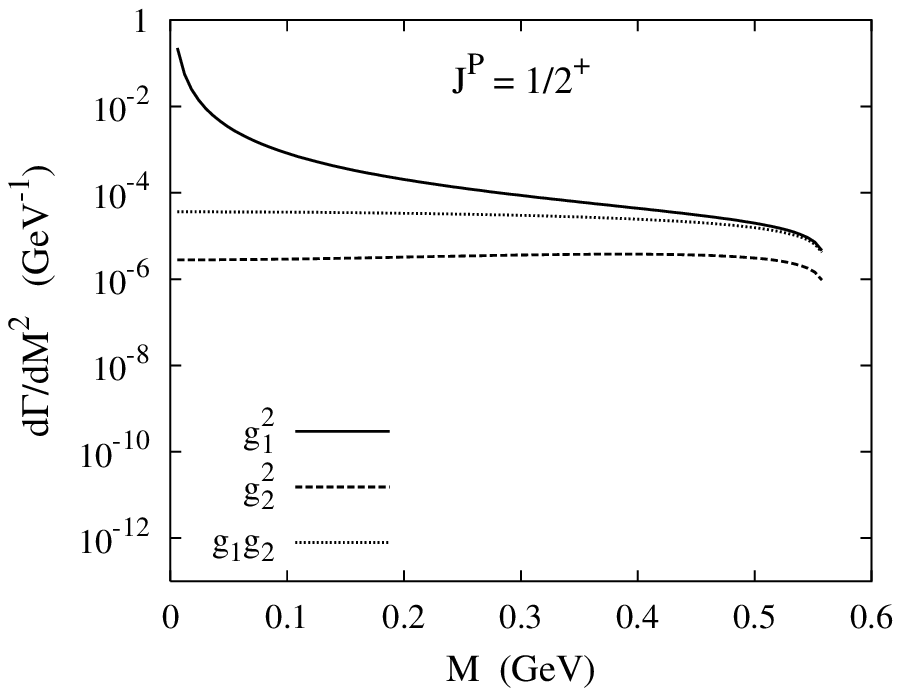}
\includegraphics[width=6.2cm]{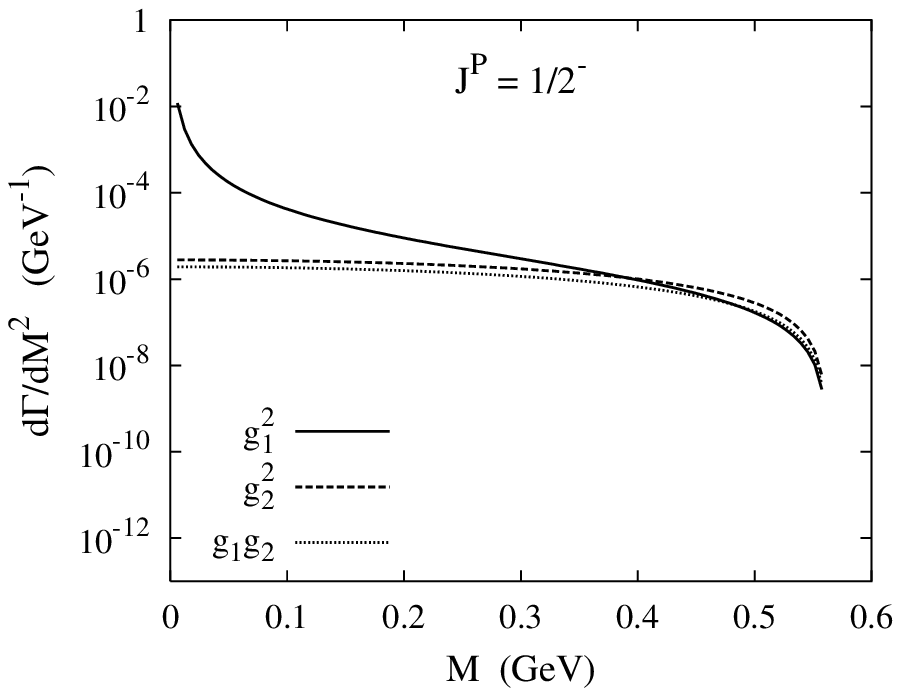}\\
\includegraphics[width=6.2cm]{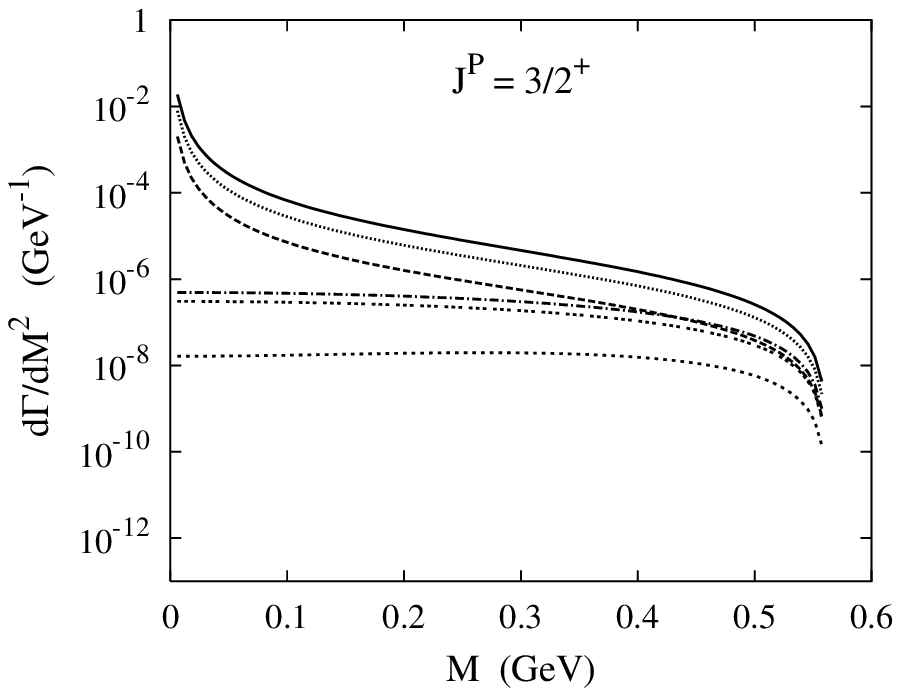}
\includegraphics[width=6.2cm]{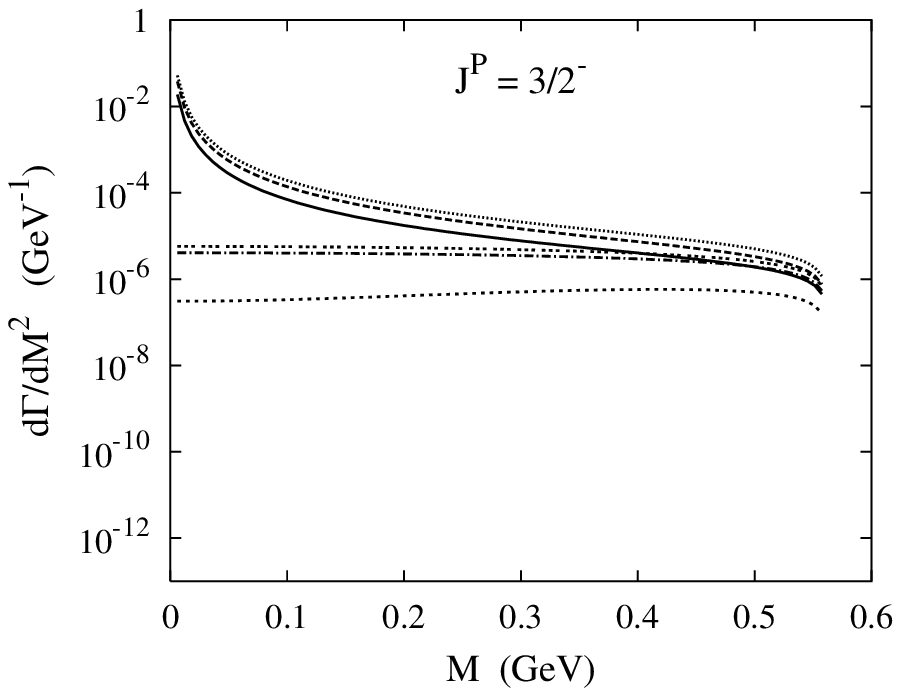}\\
\includegraphics[width=6.2cm]{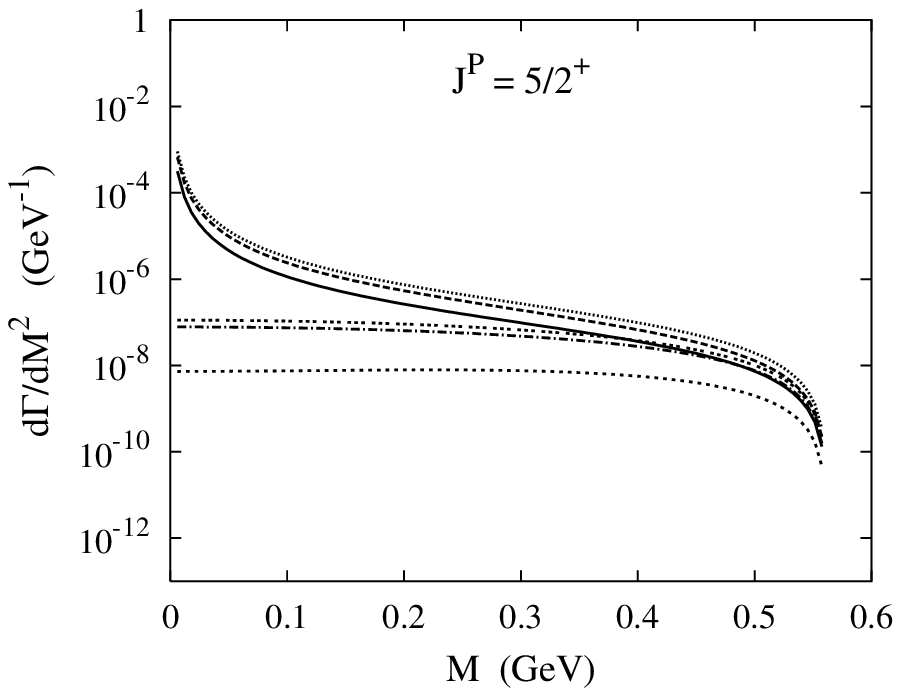}
\includegraphics[width=6.2cm]{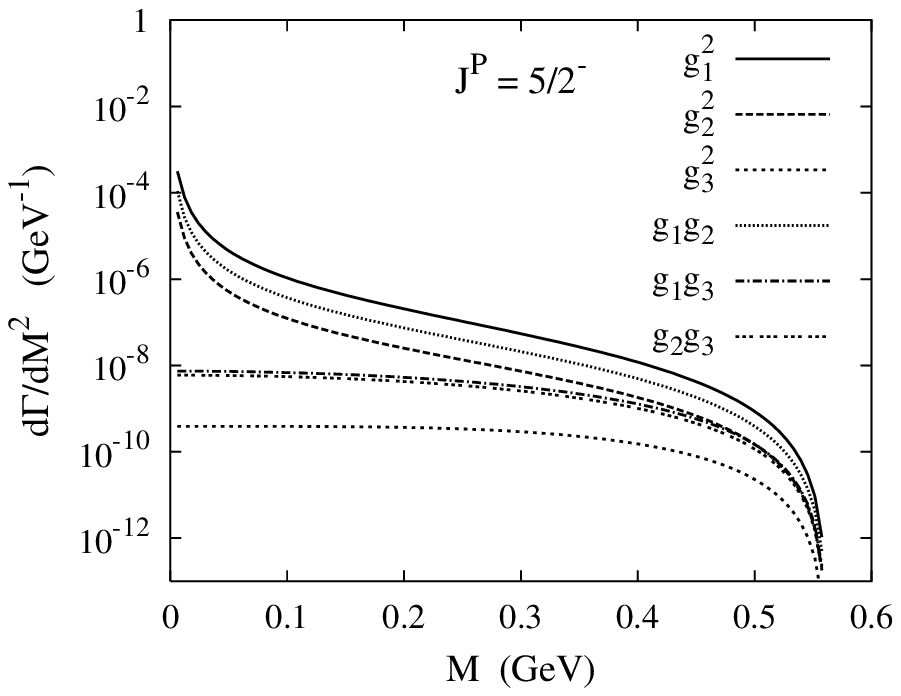}
\end{center}
\caption{Contributions of the various terms to the 
differential width of the Dalitz decay $R \to N e^+ e^-$
of an $R$ resonance with mass $m_*$ = 1.5 GeV and with various
spin-parities. The dimensionless coupling constants are set to unity.
The explanation of the linestyles in the Figure for $J^P$ = 5/2$-$
is valid for all plots for $J\ge$3/2.}
\label{dalwidfig}
\end{figure}

We first compare the contributions of the various terms in the matrix 
elements (\ref{mxelement}) and (\ref{mxelement1/2}) to the dilepton 
invariant mass spectrum. For this purpose we have set the dimensionless
couplings $g_i$ to unity and calculated the contributions of the 
individual terms and the interference terms to the differential 
Dalitz decay width. We carried out these calculations for 
spin$\le$5/2 resonances and chose a resonance mass of 1.5 GeV. 
The resulting plots are shown in Fig.\ \ref{dalwidfig}

We notice that the contribution of terms containing $g_2$ in the spin-1/2
case and terms containing $g_3$ in the spin$\ge$3/2 case
is free from singularities at $M^2 = 0$.
The real photonic width of the resonance 
$\Gamma_{R \to N\gamma}(M=0)$ obtained from these terms is zero
in accordance with Eq.\ (\ref{dalwid}), consequently, the real photonic
decay width cannot be used to fix the coupling
constants of these terms.
The Dalitz decay width calculated from these terms is significantly
smaller than the other contributions throughout the spectrum 
(with the exception of the 1/2$-$ case at large dilepton masses), 
therefore they would need a much larger coupling constant 
to give a comparable contribution.
There seems to be no natural explanation for such a huge difference
of the coupling constants.
In the following we discard the terms discussed above and make the 
assumption that in the spin-1/2 case only the $g_1$ term and in the 
spin$\ge$3/2 case only the $g_1$ and $g_2$ terms contribute to the
Dalitz decay width.

\begin{figure}[p]
\begin{center}
\includegraphics[width=6.2cm]{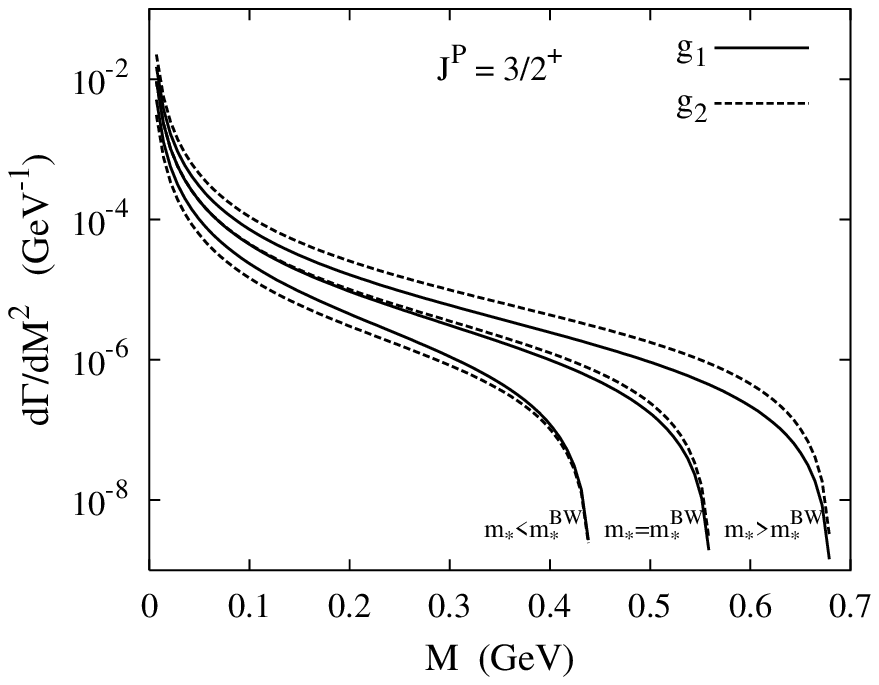}
\includegraphics[width=6.2cm]{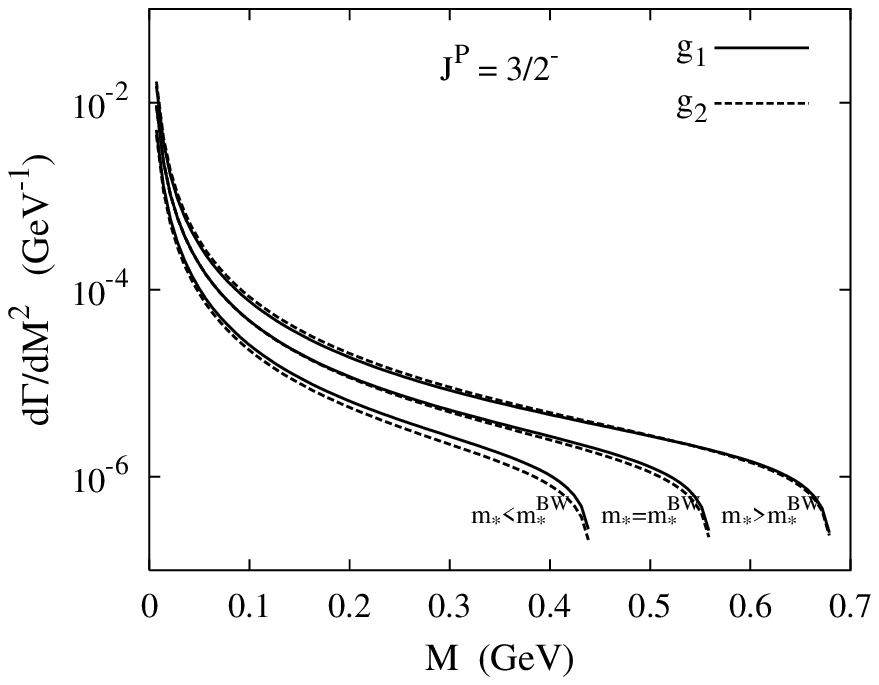}\\
\includegraphics[width=6.2cm]{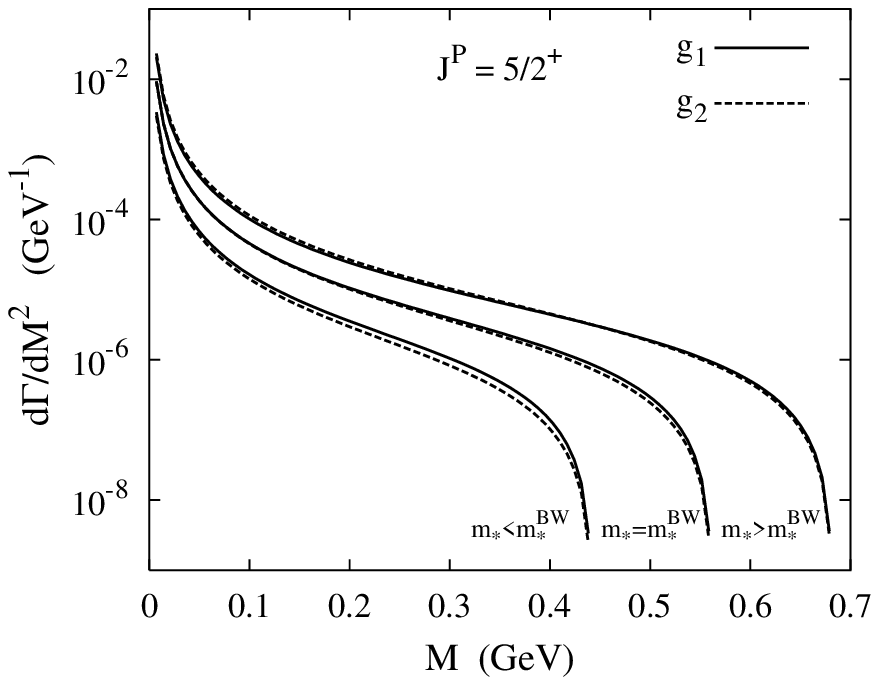}
\includegraphics[width=6.2cm]{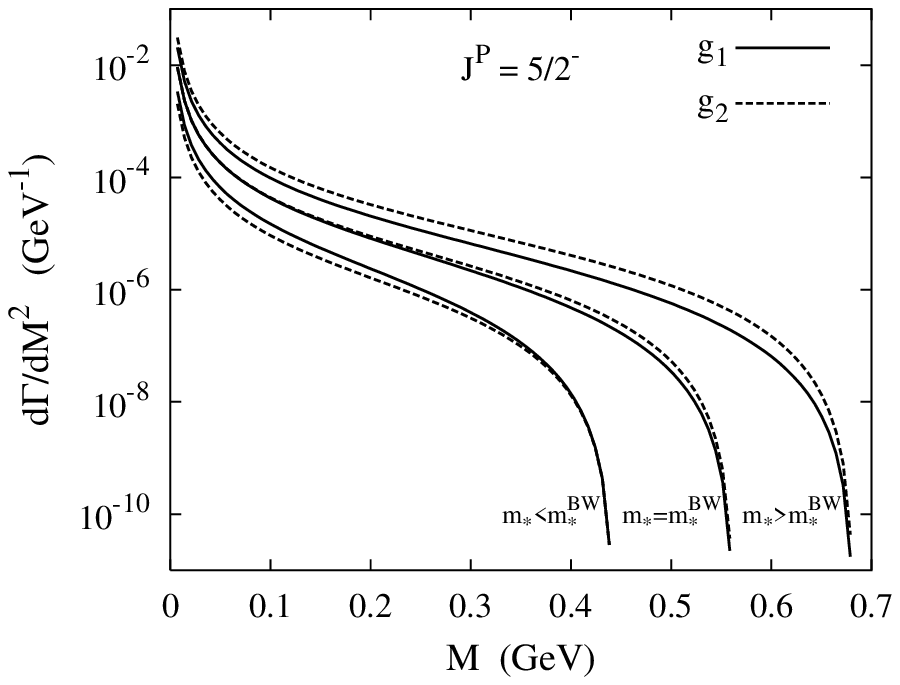}
\end{center}
\caption{Differential Dalitz decay width of a hypothetical baryon resonance 
with Breit-Wigner mass $m_*^{BW}$ = 1.5 GeV, full width 
$\Gamma_{tot}$ = 0.12 GeV and a photonic branching ratio of 5\%.
The two linestyles correspond to the assumptions that only the 
$g_1$ or $g_2$ terms contribute ($g2\approx 0$ or $g1\approx 0$).
Predictions for resonances with
masses above and below the Breit-Wigner mass 
($m_* = m_*^{BW} \pm \Gamma_{tot}$) are also shown.}
\label{mrdep}
\end{figure}

\begin{figure}[t]
\begin{center}
\includegraphics[width=7cm]{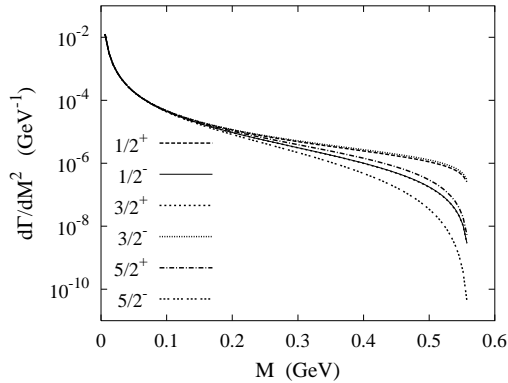}
\end{center}
\caption{Differential Dalitz decay width of hypothetical baryon
resonances with mass $m_*$ = 1.5 GeV, photonic width
$\Gamma_{R\to N\gamma}$ = 0.006 GeV and with various spin-parities.}
\label{spinpar}
\end{figure}

In Fig.\ \ref{mrdep} the differential Dalitz decay width of hypothetical 
resonances with Breit-Wigner mass $m_*^{BW}$ = 1.5 GeV, 
full width $\Gamma_{tot}$ = 0.12 GeV
and a photonic branching ratio of 5\% is shown for various spin-parities
(spin$\ge$3/2). The two linestyles refer to the results obtained
assuming that only the $g_1$ or $g_2$ term contributes ($g_2\approx 0$
or $g_1\approx 0$).
We also show the results for resonances with considerably larger and smaller
mass, specifically, $m_* = m_*^{BW} + \Gamma_{tot}$ and 
$m_* = m_*^{BW} - \Gamma_{tot}$. The real photonic width was used to fix
the coupling constants in the $m_* = m_*^{BW}$ case
and the same coupling constants have been used in the 
$m_* = m_*^{BW} \pm \Gamma_{tot}$ cases.

For $m_* = m_*^{BW}$ the $g_1$ and $g_2$ curves nearly coincide
showing that these contributions are practically indistinguishable in 
the dilepton invariant mass spectrum if the mass of the decaying
resonance is close to the Breit-Wigner mass.
However, in the spin-parity 3/2+ and 5/2$-$ cases the difference between 
the $g_1$ and $g_2$ contributions is sizable if the resonance mass is well
above the Breit-Wigner mass.
Since in transport models masses of resonances are usually not far from 
the Breit-Wigner mass, transition from one of the $g_1$ and $g_2$ terms 
to the other or to a linear combination will not have a substantial effect 
on the predictions for the dilepton invariant mass spectrum.
In the following we make the assumption that only the $g_1$ term 
contributes to the Dalitz decay width of baryon resonances.
The explicite expressions for the polarization averaged squared transition
matrix elements calculated from the $g_1$ terms are given in Appendix B.

To demonstrate the significance of spin and parity we show in Fig.\ 
\ref{spinpar} the differential Dalitz decay width of hypothetical
resonances with mass $m_*$ = 1.5 GeV, photonic decay width
$\Gamma_{R \to N\gamma}$ = 0.006 GeV and with various spin-parities.
The invariant mass spectra show large differences in the large dilepton 
mass range. The largest contribution comes from spin-parity 1/2+
and 3/2$-$ resonances.

\begin{figure}[p]
\begin{center}
\includegraphics[width=6.2cm]{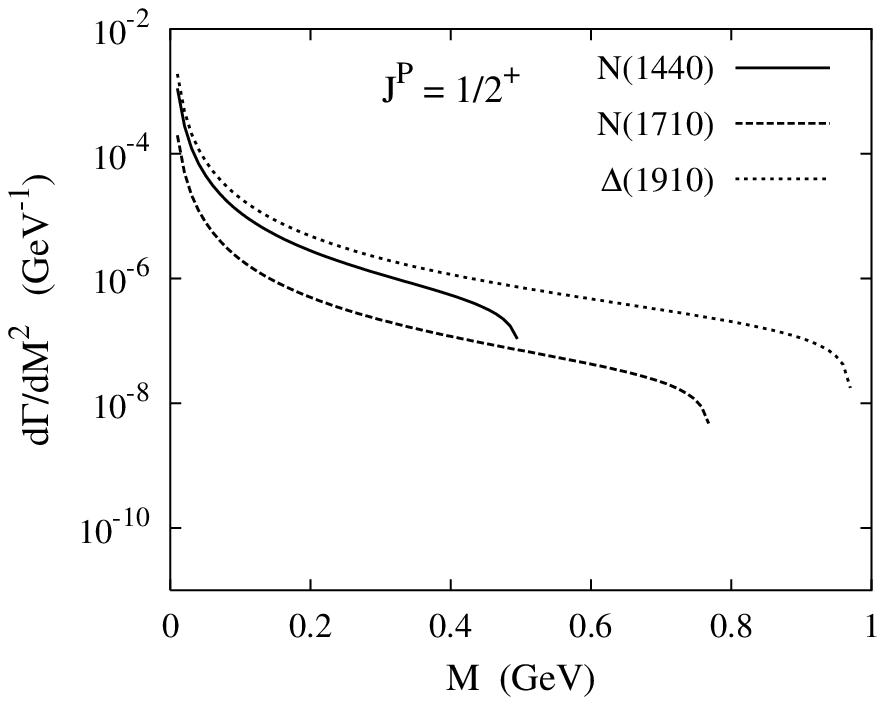}
\includegraphics[width=6.2cm]{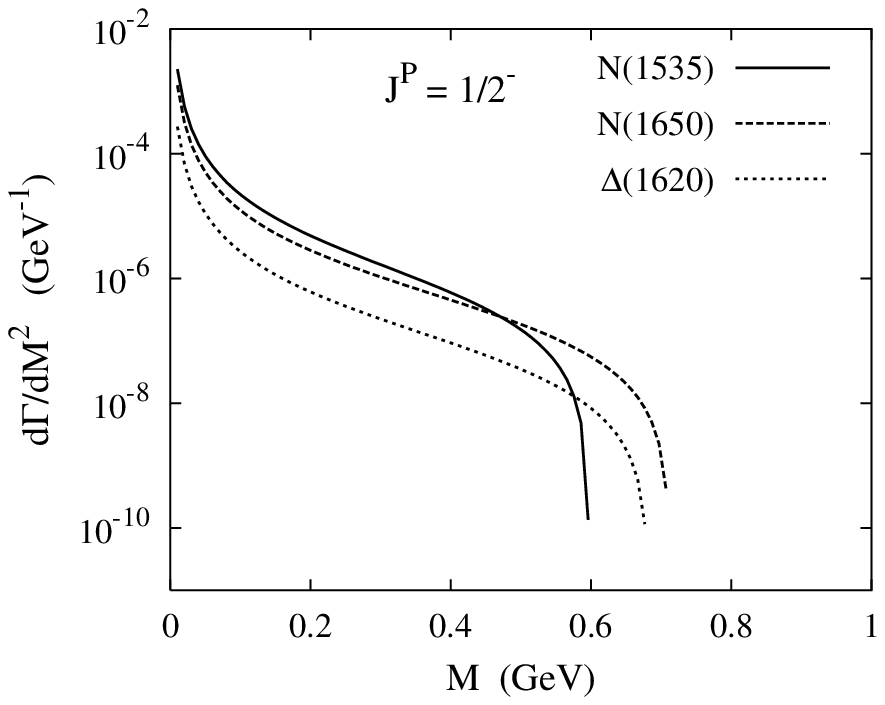}\\
\includegraphics[width=6.2cm]{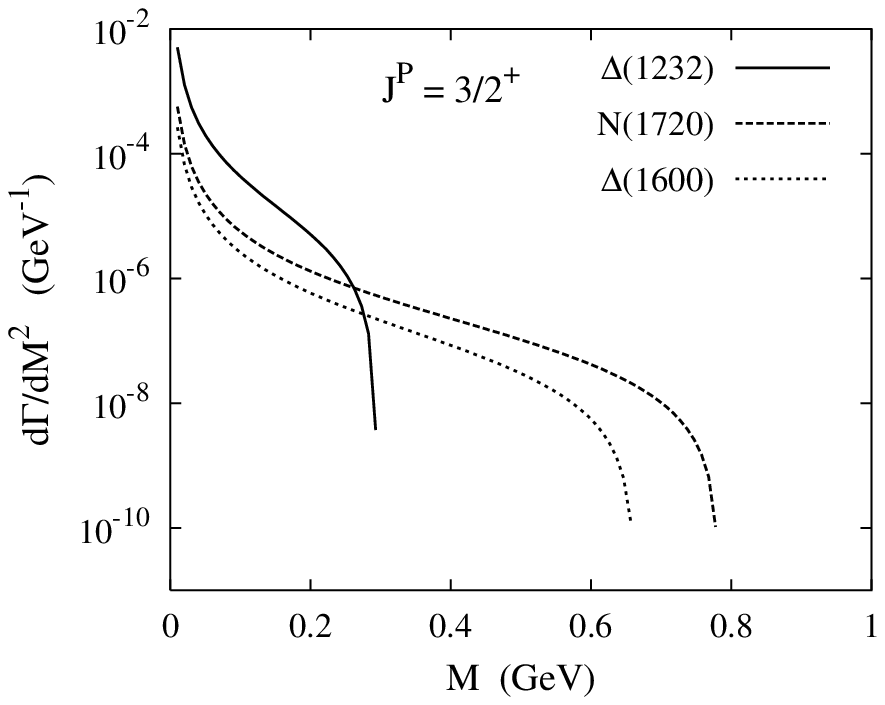}
\includegraphics[width=6.2cm]{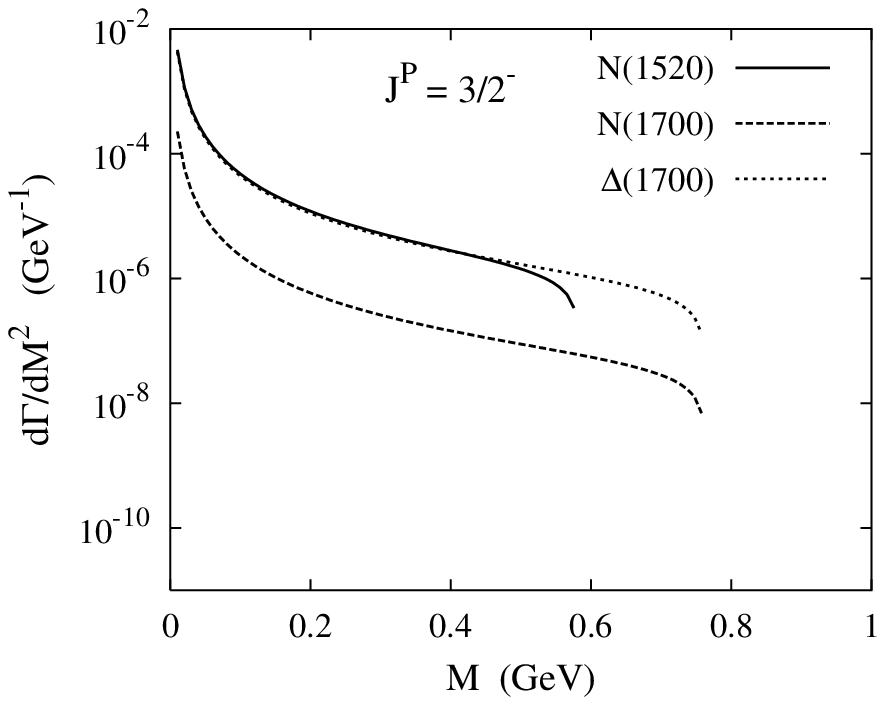}\\
\includegraphics[width=6.2cm]{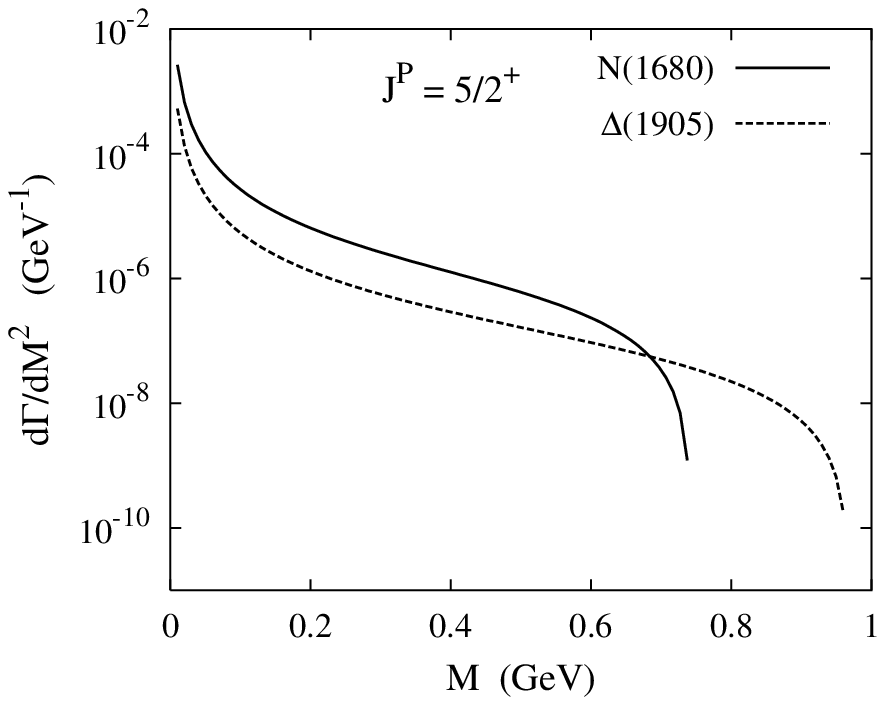}
\includegraphics[width=6.2cm]{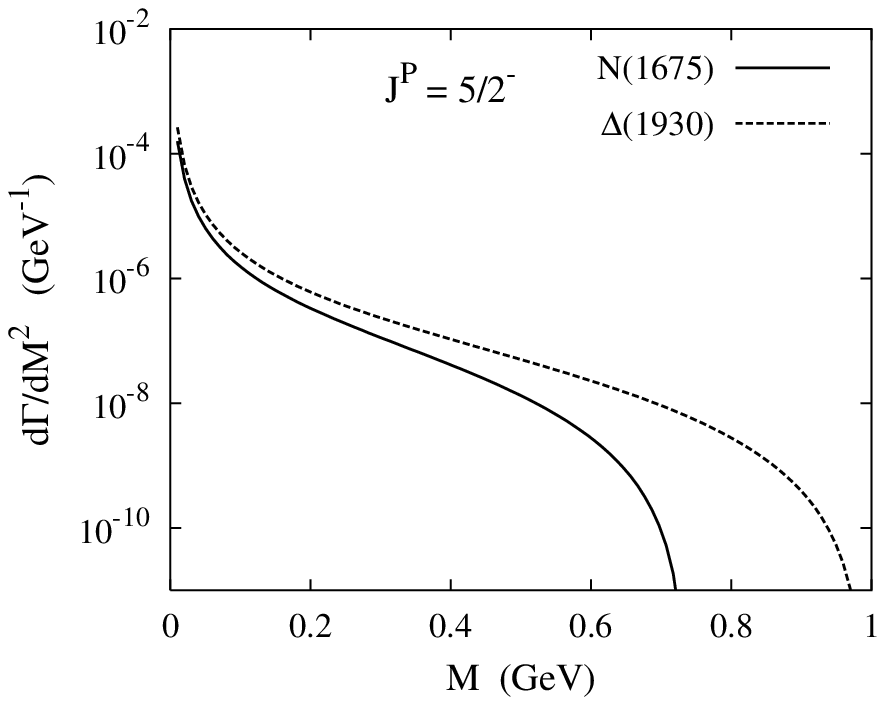}
\end{center}
\caption{Differential Dalitz decay width of unflavored baryon resonances.
The curves correspond to resonances with $m_* = m_*^{BW}$.
The Dalitz decay width of resonances with a mass different from the 
Breit-Wigner mass can be significantly different (c.f.\ Fig.\ \ref{mrdep}).}
\label{real}
\end{figure}

\begin{table}[p]
\begin{center}
\begin{tabular}{c|c|l|cc|cc|}
$J^P$ & resonance & $\Gamma_{tot}$ & 
\multicolumn{2}{c|}{Branching ratios (\%)} &
 \multicolumn{2}{c|}{Coupling constants} \\
 &  & & $p\gamma$ & $n\gamma$ & $g_1^{p\gamma}$ & $g_1^{n\gamma}$ \\
\hline
\hline
& & & & & & \\[-10pt]
1/2+ & N(1440) & 0.35 & 0.042 & 0.021 & 0.139 & 0.098 \\
     & N(1710) & 0.1 & 0.026 & 0.01{\phantom{$0$}} & 0.030 & 0.019 \\
     & $\Delta$(1910) & 0.25 & \multicolumn{2}{c|}{0.1{\phantom{$00$}}} & 
\multicolumn{2}{c|}{0.066}  \\
& & & & & & \\[-10pt]
\hline
& & & & & & \\[-10pt]
1/2$-$ & N(1535) & 0.15 & 0.2{\phantom{$00$}} & 0.15{\phantom{$0$}} & 
0.634 & 0.549 \\
     & N(1650) & 0.15 & 0.11{\phantom{$0$}} & 0.09{\phantom{$0$}} & 
0.315 & 0.285 \\
     & $\Delta$(1620) & 0.15 & \multicolumn{2}{c|}{0.024} & 
\multicolumn{2}{c|}{0.162} \\
& & & & & & \\[-10pt]
\hline
& & & & & & \\[-10pt]
3/2+ & $\Delta$(1232) & 0.12 & \multicolumn{2}{c|}{0.56{\phantom{$0$}}} & 
\multicolumn{2}{c|}{1.98{\phantom{$0$}}} \\
     & N(1720) & 0.15 & 0.05{\phantom{$0$}} & 0.2{\phantom{$00$}} & 
0.193 & 0.386 \\
     & $\Delta$(1600) & 0.35 & \multicolumn{2}{c|}{0.01{\phantom{$0$}}} & 
\multicolumn{2}{c|}{0.162} \\
& & & & & & \\[-10pt]
\hline
& & & & & & \\[-10pt]
3/2$-$ & N(1520) & 0.12 & 0.51{\phantom{$0$}} & 0.42{\phantom{$0$}} & 
0.793 & 0.719 \\
     & N(1700) & 0.1 & 0.03{\phantom{$0$}} & 0.07{\phantom{$0$}} & 
0.126 & 0.193 \\
     & $\Delta$(1700) & 0.3 & \multicolumn{2}{c|}{0.19{\phantom{$0$}}} & 
\multicolumn{2}{c|}{0.549} \\
& & & & & & \\[-10pt]
\hline
& & & & & & \\[-10pt]
5/2+ & N(1680) & 0.13 & 0.27{\phantom{$0$}} & 0.034 & 2.74{\phantom{$0$}} & 
0.971 \\
     & $\Delta$(1905) & 0.35 & \multicolumn{2}{c|}{0.02{\phantom{$0$}}} & 
\multicolumn{2}{c|}{0.713} \\
& & & & & & \\[-10pt]
\hline
& & & & & & \\[-10pt]
5/2$-$ & N(1675) & 0.15 & 0.014 & 0.07{\phantom{$0$}} & 0.678 & 
1.52{\phantom{$0$}} \\
     & $\Delta$(1930) & 0.35 & \multicolumn{2}{c|}{0.01{\phantom{$0$}}} 
& \multicolumn{2}{c|}{0.479} \\[-10pt]
& & & & & & \\
\hline
\hline
\end{tabular}
\end{center}
\vspace*{-14pt}
\caption{Coupling constants $g_1$ for the electromagnetic transitions
of baryon resonances. In the case of nucleon resonances we give separately
the coupling constants for the $p\gamma$ and $n\gamma$ channels.
We also give the full widths and photonic branching ratios used to
obtain the values of the coupling constants.}
\label{couplings}
\end{table}

The differential Dalitz decay width of real baryon resonances is shown
in Fig.\ \ref{real}. We included all spin$\le$5/2 nucleon and $\Delta$
resonances below 2 GeV listed in the Review of Particle Physics 
\cite{PDG} with at least *** status. We 
assumed that $g_2 = g_3 = 0$ and fitted
the $g_1$ coupling constants to the $R^+ \to p\gamma$ decay width,
the numerical values of 
which we obtained using the Breit-Wigner full width and the $p\gamma$
branching ratio of the resonances given in \cite{PDG}. For the branching ratio,
which is poorly known in many cases, we used the average of the lower 
and upper bounds. In the case of $\Delta$ resonances, where the $p\gamma$ 
and $n\gamma$ branching ratios are not listed separately, 
we used the $N\gamma$ partial width to fix the coupling constants. 
Note that the $p\gamma$ and$n\gamma$ branching ratios may differ significantly.

In Table \ref{couplings} we give the resulting coupling constants
together with the values of the full width and the photonic branching ratio
used in the fit. In the case of nucleon resonances we also give
the $n\gamma$ couplings.

%%%%%%%%%%%%%%%%%%%%%%%%%%%%%%%%%%%%%%%%%%%%%%%%%%%%%%%%%%%%%%%%%%%%%%%%

\section{Conclusion}

In this paper we studied the Dalitz decay of baryon resonances.
We have given expressions for the differential Dalitz decay width of
resonances with spin$\le$5/2. The number of independent terms
in the electromagnetic transition matrix element is 2 in the
spin-1/2 case and 3 in the spin$\ge$3/2 case. One of these terms
does not contribute to the real photonic decay of the resonances,
which is used to fix the coupling constants.
We have shown that the remaining two matrix elements in the spin$\ge$3/2
case provide very similar contributions to the dilepton invariant mass
spectrum. On the other hand, expressions for different spin-parities give 
very different dilepton spectra in the large dilepton mass region.
Our results can be used in models of dielectron production in elementary
hadronic reactions or heavy ion collisions.

%%%%%%%%%%%%%%%%%%%%%%%%%%%%%%%%%%%%%%%%%%%%%%%%%%%%%%%%%%%%%%%%%%%%%

\section*{Acknowledgments}

This work was supported by the National Fund for Scientific Research
of Hungary, grant Nos.\ OTKA T30171, T30855 and T32038.

%%%%%%%%%%%%%%%%%%%%%%%%%%%%%%%%%%%%%%%%%%%%%%%%%%%%%%%%%%%%%%%%%%%%%%

\appendix
\section*{Appendix A}

We give here the spin projectors needed to calculate
the square of the transition matrix elements. 
Throughout the paper we use the conventions of \cite{BD}
for relativistic quantities and Dirac matrices and spinors.
Specifically, we use the normalization
$\bar{u}(p,s)u(p,s) = 2m$ for Dirac spinors ($p^2 = m^2$),
therefore
\be
\sum_{s}u(p,s)\bar{u}(p,s) = \psl + m.
\ee
The spin-3/2 projector:
\be
\sum_{s}u^{\mu}(p,s)\bar{u}^{\nu}(p,s) =
(\psl + m)
\left[g^{\mu\nu} - \frac{\gamma^{\mu}\gamma^{\nu}}{3}
      - \frac{2}{3}\frac{p^{\mu}p^{\nu}}{m^2}
      + \frac{p^{\mu}\gamma^{\nu} - p^{\nu}\gamma^{\mu}}{3m}
\right].
\ee
The spin-5/2 projector:
\bea
\lefteqn{\sum_{s}u^{\mu\nu}(p,s)\bar{u}^{\rho\sigma}(p,s) =
(\psl + m)} \nn \\
 & & \times
\left[ \frac{3}{10}\left(G^{\mu\rho}G^{\nu\sigma} 
                       + G^{\mu\sigma}G^{\nu\rho}
                   \right)
     - \frac{1}{5}G^{\mu\nu}G^{\rho\sigma} \right. \nn \\
 & & \left. - \frac{1}{10}\left(T^{\mu\rho}G^{\nu\sigma}
                       + T^{\nu\sigma}G^{\mu\rho}
                       + T^{\mu\sigma}G^{\nu\rho}
                       + T^{\nu\rho}G^{\mu\sigma}
                   \right)
\right],
\eea
with
\be
G^{\mu\nu} = - g^{\mu\nu} + \frac{p^{\mu}p^{\nu}}{m^2},
\ee
and
\be
T^{\mu\nu} = 
- \frac{1}{2}(\gamma^{\mu}\gamma^{\nu}-\gamma^{\nu}\gamma^{\mu})
+ \frac{p^{\mu}\left(\psl\gamma^{\nu}
                   - \gamma^{\nu}\psl
               \right)}{2m^2}
- \frac{p^{\nu}\left(\psl\gamma^{\mu}
                   - \gamma^{\mu}\psl
               \right)}{2m^2}.
\ee

\section*{Appendix B}

In Eq.\ (\ref{photwid})
the squared transition matrix element averaged over initial and summed over
final polarization states can be written as
\be
\frac{1}{n_{pol,R}}
\sum_{pol} {\vert \langle N \gamma | T | R \rangle \vert}^2 =
\left( -g^{\mu\nu} + \frac{q^{\mu}q^{\nu}}{q^2} \right) 
{\mathcal M}_{\mu\nu},
\ee
where the notation
\be 
{\mathcal M}_{\mu\nu} =
\frac{1}{n_{pol,R}}
\sum_{pol_{N,R}} \langle N | J_{\mu} | R \rangle
 \langle R | J_{\nu} | N \rangle
\ee
was used.
Following \cite{LS} we introduce the notations
\bea
L_{\mu} & = & (p_*\cdot q)q_{\mu} - q^2 p_{*\mu}, \nn \\
L_{\mu\nu} & = & \frac{L_{\mu}L_{\nu}}{L^2}, \nn \\
T_{\mu\nu} & = & g_{\mu\nu} - \frac{q^{\mu}q^{\nu}}{q^2} - L_{\mu\nu}.
\eea
${\mathcal M}_{\mu\nu}$ can be decomposed in terms of transverse
and longitudinal parts, ${\mathcal M}_T$ and ${\mathcal M}_L$,
in the form
\be
{\mathcal M}_{\mu\nu} = - {\mathcal M}_T T_{\mu\nu}
 - {\mathcal M}_L L_{\mu\nu}.
\ee
In terms of ${\mathcal M}_T$ and ${\mathcal M}_L$ the polarization
averaged squared transition matrix element is expressed as
\be
\frac{1}{n_{pol,R}}
\sum_{pol} {\vert \langle N \gamma | T | R \rangle \vert}^2 =
2 {\mathcal M}_T + {\mathcal M}_L.
\ee
Here we give the contributions of the $g_1^2$ terms to 
${\mathcal M}_T$ and ${\mathcal M}_L$ for the various spin-parity cases:
\bea
{\mathcal M}_T^{1/2+} & = & 4\pi\alpha g_1^2
\frac{1}{2m^4}\left(m_*^2-m^2\right)^2\left[(m_*+m)^2-M^2\right], \\
{\mathcal M}_L^{1/2+} & = & 4\pi\alpha g_1^2
\frac{M^2}{2m^4}(m_*+m)^2\left[(m_*+m)^2-M^2\right], \\
{\mathcal M}_T^{1/2-} & = & 4\pi\alpha g_1^2
\frac{1}{2m^4}\left(m_*^2-m^2\right)^2\left[(m_*-m)^2-M^2\right], \\
{\mathcal M}_L^{1/2-} & = & 4\pi\alpha g_1^2
\frac{M^2}{2m^4}(m_*-m)^2\left[(m_*-m)^2-M^2\right], \\
{\mathcal M}_T^{3/2+} & = & 4\pi\alpha g_1^2
\frac{1}{12 m_*^2 m^2}\left[(m_*-m)^2-M^2\right] \nn \\
 & & \times \left(3m_*^4 + 6m_*^3m + 4m_*^2m^2 + 2m_*m^3 + m^4 \right. \nn \\
 & & \left. - 2m_*mM^2 - 2m^2M^2 + M^4\right), \\
{\mathcal M}_L^{3/2+} & = & 4\pi\alpha g_1^2
\frac{M^2}{3m^2}\left[(m_*-m)^2-M^2\right], \\
{\mathcal M}_T^{3/2-} & = & 4\pi\alpha g_1^2
\frac{1}{12 m_*^2 m^2}\left[(m_*+m)^2-M^2\right] \nn \\
 & & \times \left(3m_*^4 - 6m_*^3m + 4m_*^2m^2 - 2m_*m^3 + m^4 \right. \nn \\
 & & \left. + 2m_*mM^2 - 2m^2M^2 + M^4\right), \\
{\mathcal M}_L^{3/2-} & = & 4\pi\alpha g_1^2
\frac{M^2}{3m^2}\left[(m_*+m)^2-M^2\right], \\
{\mathcal M}_T^{5/2+} & = & 4\pi\alpha g_1^2
\frac{1}{480 m_*^4 m^4}\left[(m_*-m)^2-M^2\right]
 \left[(m_*+m)^2-M^2\right]^2 \nn \\
 & & \times \left(2m_*^4 - 4m_*^3m + 3m_*^2m^2 - 2m_*m^3 + m^4 \right. \nn \\
 & & \left. + 2m_*mM^2 - 2m^2M^2 + M^4\right), \\
{\mathcal M}_L^{5/2+} & = & 4\pi\alpha g_1^2
\frac{M^2}{120m_*^2m^4}\left[(m_*-m)^2-M^2\right]
 \left[(m_*+m)^2-M^2\right]^2, \\
{\mathcal M}_T^{5/2-} & = & 4\pi\alpha g_1^2
\frac{1}{480 m_*^4 m^4}\left[(m_*-m)^2-M^2\right]^2
 \left[(m_*+m)^2-M^2\right] \nn \\
 & & \times \left(2m_*^4 + 4m_*^3m + 3m_*^2m^2 + 2m_*m^3 + m^4 \right. \nn \\
 & & \left. - 2m_*mM^2 - 2m^2M^2 + M^4\right), \\
{\mathcal M}_L^{5/2-} & = & 4\pi\alpha g_1^2
\frac{M^2}{120m_*^2m^4}\left[(m_*-m)^2-M^2\right]^2
 \left[(m_*+m)^2-M^2\right].
\eea

%%%%%%%%%%%%%%%%%%%%%%%%%%%%%%%%%%%%%%%%%%%%%%%%%%%%%%%%%%%%%%%%%%%%

\end{document}